# On the Design of Universal LDPC Codes

Ali Sanaei, Mahdi Ramezani, and Masoud Ardakani

Department of Electrical and Computer Engineering, University of Alberta, Canada
Email: {sanaei,ramezani,ardakani}@ece.ualberta.ca

*Abstract*—Low-density parity-check (LDPC) coding for a multitude of equal-capacity channels is studied. First, based on numerous observations, a conjecture is stated that when the belief propagation decoder converges on a set of equal-capacity channels, it would also converge on any convex combination of those channels. Then, it is proved that when the stability condition is satisfied for a number of channels, it is also satisfied for any channel in their convex hull. For the purpose of code design, a method is proposed which can decompose every symmetric channel with capacity $\mathcal{C}$ into a set of identical-capacity basis channels. We expect codes that work on the basis channels to be suitable for any channel with capacity $\mathcal{C}$. Such codes are found and in comparison with existing LDPC codes that are designed for specific channels, our codes obtain considerable coding gains when used across a multitude of channels.

## I. Introduction

Design of codes for specific channels is a mature subject. A code optimized for one channel, however, may not perform well if used on other channel types [1]. Thus, recently, there has been an increasing interest in universal codes, i.e., codes that perform well on a multitude of channels [2]–[4]. Such codes reduce system complexity by removing a need for frequent code changes in the system and by allowing for once-and-for-all coding solutions.

Low-density parity-check (LDPC) codes [5] are extremely strong error correcting codes. Interestingly, various authors have observed "universal properties" of these codes [3], [4], [6]–[8]. Chung [6] points out that LDPC codes optimized for the Gaussian channel perform well on some other channels such as the Rayleigh channel. In a more general setup, Shi and Wesel [4] discuss the universal properties of finite block length codes.

Peng *et al.* [8] design LDPC codes for a number of channels (in this case, the Gaussian channel, the binary erasure channel (BEC) and the Rayleigh channel). They argue that for a set of channels, usually one channel can be taken as the surrogate. They design the code only for the surrogate channel. This code works satisfactorily on all the given channels, but not necessarily on other channel types.

Despite some universal properties of LDPC codes, the performance of a code designed for one channel, can be poor on another channel with the same capacity. For example, a rate one-half LDPC code with maximum node degree of 100 (taken from [9]), achieving more than $99.7\%$ of the capacity of a BEC with capacity $\mathcal{C} = 0.5$, does not converge on a binary symmetric channel (BSC) with capacity $\mathcal{C} = 0.63$. In other words, this code does not achieve even $80\%$ of the capacity of the BSC.

In this work, we design LDPC codes that have strong universal properties. For a given channel capacity $\mathcal{C}$, we find LDPC codes that perform well on any channel with this capacity. The main body of this paper consists of: (1) studying code design and stability analysis for convex combination of $N$ channels of equal capacity, and (2) decomposition of all channels with capacity $\mathcal{C}$ into a number of basis channels of the same capacity.

In Section III, we study code design for the convex combinations of a set of equal-capacity channels (see Section III for definition). We conjecture that under belief propagation decoding, to design a code for all the convex combinations of $N$ binary-input symmetric-output (BISO) channels, it is sufficient to find a code which converges only on those $N$ channels. We also prove that when stability condition [10] is satisfied for $N$ BISO channels, it is also satisfied for all their convex combinations. As a result, if a set of basis channels with capacity $\mathcal{C}$ can be found, a code designed only for the basis channels has strong universal properties.

In Section IV, a channel decomposition method is suggested which decomposes any BISO channel with capacity $\mathcal{C}$ in terms of basis channels of the same capacity with nonnegative coefficients. This decomposition method is not similar to existing techniques that decompose the channel over BSCs of various capacities [11]. Our technique is exhaustive, i.e., any channel with a given capacity can be spanned with nonnegative coefficients over our suggested basis.

For any given capacity $\mathcal{C}$, a code designed only for our suggested basis channels is expected to have strong universal behavior over all channels with capacity $\mathcal{C}$. Examples are provided in Section V. Simulations confirm that compared to codes designed for specific channels, our codes have significantly better universal properties.

## II. Preliminaries: LDPC Code Design and Symmetric Channel Representations

In this paper, an ensemble of LDPC codes is defined by a pair of distributions $(\lambda, \rho)$ in the polynomial form, i.e., $\lambda(x) = \sum_{i \geq 2} \lambda_i x^{i-1}$ and $\rho(x) = \sum_{i \geq 2} \rho_i x^{i-1}$, where $\lambda_i$ ($\rho_i$) is the fraction of edges connected to the variable (check) nodes of degree $i$.

LDPC code design means optimizing degree distributions in order to maximize a cost function which can be usually the code threshold or the code rate. In threshold maximization given a fixed rate, we seek a code working under the worst channel condition, hence minimum gap between the given rate and channel capacity. In rate maximization given a fixed channel, the code which exhibits maximum rate and provides reliable communication over the given channel is found.

In this work, we focus on finding codes suitable for a multitude of channels rather than a single channel. We adopt the rate maximization approach which allows us to find a code working on a finite number of channels. Using the rate maximization approach, code design for a number of identical-capacity channels is similar to what has been discussed in [12].

We assume that transmission of LDPC codes takes place on a memoryless BISO channel, i.e., $(\mathcal{X}, W(Y|X), \mathcal{Y})$ where the input alphabet $\mathcal{X} = \{0, 1\}$ is uniformly distributed, $W(Y|X)$ is the conditional pdf of the channel and $\mathcal{Y} \subseteq \mathbb{R}$ is the output alphabet of the channel. In this section, three channel representations for a BISO channel are presented. A BISO channel is a channel for which we have

$$W(Y|X=1) = W(-Y|X=0).$$

At the receiver side, given the output $Y$, the sufficient statistics for $X$ is $\Pr(X=0|Y)$ which depends only on the conditional pdf of the channel. For a symmetric channel, the log-likelihood ratio (LLR) is defined as

$$M = \log \frac{\Pr(X=0|Y)}{\Pr(X=1|Y)}.$$

It is straightforward to see that $M$ is another sufficient statistics for $X$, i.e., $I(X;Y) = I(X;M)$. Assuming that the all-zero codeword is transmitted, we denote the pdf of $M$ as $\mathrm{a}(x)$ which fully characterizes the channel and satisfies the symmetric property $\mathrm{a}(-x) = e^{-x}\mathrm{a}(x)$ [10]. Since the channel is symmetric and input symbols are equiprobable, the mutual information is the same as capacity [13]. We denote the mutual information $I(X;Y)$ associated with $\mathrm{a}(x)$ by $\mathcal{I}(\mathrm{a})$[1] defined as

$$\mathcal{I}(\mathrm{a}) = 1 - \int \mathrm{a}(x) \log_2(1 + e^{-x}) dx.$$

Let

$$P = \min\left\{\Pr(X=0|Y), \Pr(X=1|Y)\right\}$$

be a random variable over $[0, \frac{1}{2}]$ with the pdf denoted by $g(p)$ which completely characterizes the channel [11]. This, in fact, is the probability of error based on observation of the output. Since the pdf of $P$, i.e., $g(p)$, should not satisfy any symmetry property, we will use this representation in Section IV.

Let $\Delta_z(x)$ be the Dirac delta function at $x = z$. Also, for an $\epsilon \in [0, 1]$, define $\bar{\epsilon} = 1 - \epsilon$. A BSC with crossover probability $\epsilon$ can be represented by the pdf of LLR and the pdf of error probability as

$$\mathrm{a}(x) = \epsilon \Delta_{-\log(\frac{\bar{\epsilon}}{\epsilon})}(x) + \bar{\epsilon} \Delta_{\log(\frac{\bar{\epsilon}}{\epsilon})}(x)$$

and $g(p) = \Delta_\epsilon(p)$, respectively.

[1]For ease of notations, sometimes the argument of LLR densities is dropped.

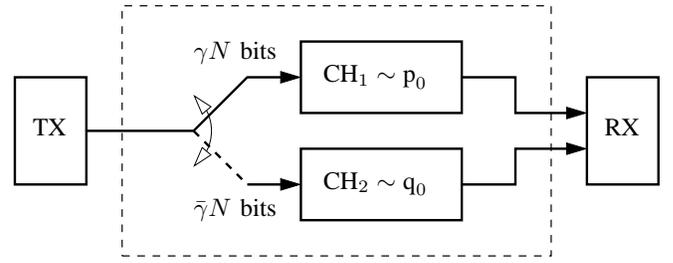

Fig. 1. Convex combination of two subchannels where each bit is transmitted through $CH_1$ with the probability of $\gamma$. The LLR densities of subchannels are $\mathrm{p}_0$ and $\mathrm{q}_0$.

Every valid pdf over $p \in [0, \frac{1}{2}]$ can be seen as a symmetric channel. In fact, this representation decomposes a symmetric channel into the convex hull of BSCs of different crossover probabilities. The fraction of bits passing through each BSC is captured in $g(p)$. According to [11], the channel capacity is

$$\mathcal{C} = \mathbb{E}\{\mathcal{C}_{\mathrm{BSC}(p)}\} = 1 - \int_0^{\frac{1}{2}} h(p)g(p)dp$$

where $h(p)$ is the binary entropy function and integration includes the mass points at $p = 0$ and $\frac{1}{2}$. Also, $g(p)$ is obtained by

$$g(p) = \frac{1}{p(1-p)}\mathrm{a}_{|m|}\left(\log\frac{1-p}{p}\right), \quad \forall p \in [0, \frac{1}{2}]$$

where $\mathrm{a}_{|m|}(x)$ is the pdf of the absolute value of LLR, i.e.,

$$\mathrm{a}_{|m|}(x) = \mathrm{a}(x)(1 + e^{-x}), \quad \forall x \geq 0.$$

The three channel representations presented in this section are interchangeable as long as the input is uniformly distributed [11].

### III. Universal Codes for Convex Combination of Two Channels and Stability Analysis

In this section, we consider universal codes for convex combination of two channels. We start with the definition of convex combination of two channels.

Fig. 1 illustrates a channel composed of two subchannels of capacity $\mathcal{C}$ and LLR densities $\mathrm{p}_0$ and $\mathrm{q}_0$. Each bit is passed through subchannel 1 with probability $\gamma \in [0, 1]$ or passed through the other subchannel with probability of $1 - \gamma$. We refer to this channel as a convex combination of subchannels 1 and 2 since the LLR density for this combination is $\gamma \mathrm{p}_0 + \bar{\gamma} \mathrm{q}_0$. The same equation holds for the pdf of error probability. We denote the set of channels in the convex hull of $\mathrm{p}_0$ and $\mathrm{q}_0$ by

$$\mathfrak{C}(\mathrm{p}_0, \mathrm{q}_0) = \{\gamma \mathrm{p}_0 + \bar{\gamma} \mathrm{q}_0 | \gamma \in [0, 1]\}.$$

Since both of the subchannels have capacity $\mathcal{C}$, for every channel $\mathrm{a}_\gamma \in \mathfrak{C}(\mathrm{p}_0, \mathrm{q}_0)$, $I(\mathrm{a}_\gamma) = \mathcal{C}$.

The setup in Fig. 1 can be viewed as an arbitrarily varying channel [14] with a Bernoulli distributed set of states. As a straightforward generalization, we can extend this configuration to $N$ subchannels.

Throughout this paper, we assume that the receiver has the channel state information and computes correct LLRs. If $\gamma$ is fixed, one can design a single good LDPC code for the channel with LLR density of $\gamma p_0 + \bar{\gamma} q_0$. However, if $\gamma$ varies over time, finding a code which works for all $\gamma \in [0, 1]$ is a challenging problem. In fact, we seek a single universal code which does not need to be changed over time, even with variations of $\gamma$.

*Remark 1:* If the transmitter knew through which of the subchannels the next transmitted bit would pass (which is not the case in most practical situations), it could use two different codes for two subchannels.

### A. Code Design for Two Channels

For code design, we use density evolution [10]. Consider a degree distribution pair $(\lambda, \rho)$ and a BISO channel with LLR density $a_0$ which can be either $p_0$ or $q_0$. Density evolution is stated as

$$a_\ell = a_0 \circledast \sum_i \lambda_i \bigg( \sum_k \rho_k a_{\ell-1}^{\boxtimes (k-1)} \bigg)^{\circledast (i-1)} \quad (1)$$

where $(\cdot)^{\circledast (i-1)}$ and $(\cdot)^{\boxtimes (k-1)}$ denote the operation of a degree-$i$ variable node and a degree-$k$ check node, respectively.

Let us first assume that there are only two possible cases of $\gamma = 0$ and $\gamma = 1$. Similar to code design for one channel, and by considering the convergence constraints on both channels, codes that are suitable for both channels can be designed. Similarly, if $\gamma$ takes only a finite number of values, the same approach can be used. While the objective function remains the same, the number of constraints grows almost linearly with the number of channels. Thus, this approach can be inefficient when $\gamma$ can take many values. In general $\gamma$ may take all values in $[0, 1]$.

Our numerous experiments show that if we guarantee convergence for two values of $\gamma$, the code will converge for all the values of $\gamma$ in between. Therefore, in order to design a code for the channel model given in Fig. 1, when $\gamma$ is unknown, we force convergence of the code only for $\gamma = 0$ and $\gamma = 1$. Our extensive experimental results have been consistent in verifying the following conjecture.

*Conjecture 1:* An LDPC code which converges over two channels with LLR pdfs $p_0$ and $q_0$, will also converge over any channel in the convex hull of those channels, i.e., over all $a_\gamma \in \mathfrak{C}(p_0, q_0)$.

*Remark 2:* The above conjecture can be proved under the assumption that the distribution of the output messages of variable nodes depends only on the information they bear and the structure of the code. According to central limit theorem, the distribution of random messages at the output of a variable node is close to a Gaussian distribution. Thus, the input distribution of a check node is a mixture of Gaussian distributions [6], [15] which is uniquely characterizes by the degree distributions and the input mutual information. It is noticeable that in this approach, we do not assume that the output distributions of the check nodes are Gaussian, which is a usual assumption in the literature [6]. Therefore, we expect codes designed based on this conjecture to have strong universal properties.

### B. Stability Analysis

Convergence analysis of an LDPC code over a symmetric channel is often performed by characterizing the fixed points of density evolution given in (1). Let

$$\mathcal{B}(a) = \int a(x) e^{-x/2} dx$$

and

$$\mathcal{P}(a) = \frac{1}{2} \int a(x) e^{-(|x/2|+x/2)} dx$$

be the Bhattacharyya constant and error probability associated with a symmetric density $a(x)$, respectively. It is noticeable that these two functionals are linear with respect to symmetric densities. Also, it is well-known that

$$2\mathcal{P}(a) \leq \mathcal{B}(a) \leq 2\big[\mathcal{P}(a)\overline{\mathcal{P}(a)}\big]^{\frac{1}{2}}. \quad (2)$$

From (1), one can see that for every symmetric channel, $\Delta_\infty$ is a fixed point of density evolution corresponding to the perfect decoding. It is desirable that the perfect decoding fixed point be stable, which means that once the probability of error gets small enough, the decoder converges to the error-free density ($\Delta_\infty$) [13]. The following theorem shows the stability of $\Delta_\infty$.

*Theorem 1:* For every channel in $\mathfrak{C}(p_0, q_0)$, once the decoder gets close to the perfect decoding, it will converge to the error-free fixed point ($\Delta_\infty$), provided that the given LDPC code converges on two subchannels $p_0$ and $q_0$.

*Proof:* Let $a_0 = [\gamma p_0 + \bar{\gamma} q_0] \in \mathfrak{C}(p_0, q_0)$ for a $\gamma \in [0, 1]$. According to [13], the Bhattacharyya functional is multiplicative under $\circledast$ operation. Also, it can be shown that for two symmetric densities u and v, we have

$$\mathcal{B}(u \boxtimes v) \leq \mathcal{B}(u) + \mathcal{B}(v) - \mathcal{B}(u)\mathcal{B}(v)$$

where the upper bound becomes tight for two BEC channels. Hence, by induction it is straightforward to see that

$$\mathcal{B}(u^{\boxtimes k}) \leq 1 - (1 - \mathcal{B}(u))^k.$$

Letting $t_\ell = \mathcal{B}(a_\ell)$, we arrive at

$$t_\ell \leq \mathcal{B}(a_0) \lambda(1 - \rho(1 - t_{\ell-1})) \quad (3)$$

which coincides with the density evolution for the BEC by substituting the inequality with an equality. Expanding (3) around zero, we get

$$t_\ell \leq \mathcal{B}(a_0) \lambda'(0) \rho'(1) t_{\ell-1} + O(t_{\ell-1}^2).$$

From the necessity part of the stability condition theorem for the belief propagation decoding [10], [16], since the given LDPC code converges over the two subchannels, we must have

$$\mathcal{B}(p_0) \lambda'(0) \rho'(1) < 1$$

and
$$\mathcal{B}(q_0)\lambda'(0)\rho'(1) < 1,$$

hence $\mathcal{B}(a_0)\lambda'(0)\rho'(1) < 1$. One can see that there exist $\mu_1 > 0$ and $\mu_2 > 0$ such that $\mathcal{B}(p_0)\lambda'(0)\rho'(1) + \mu_1 < 1$ and $\mathcal{B}(q_0)\lambda'(0)\rho'(1) + \mu_2 < 1$. Therefore,

$$\mathcal{B}(a_0)\lambda'(0)\rho'(1) + \mu < 1$$

where $\mu = \gamma\mu_1 + \bar{\gamma}\mu_2$ and $\mu > 0$. For sufficiently small $\tau > 0$ and some $\ell \in \mathbb{N}$, $t_{\ell-1} \leq \tau$, i.e., getting close enough to the perfect decoding, results in

$$t_\ell \leq (\mathcal{B}(a_0)\lambda'(0)\rho'(1) + \mu)t_{\ell-1} < t_{\ell-1}.$$

It shows that as $\ell \to \infty$, $t_\ell \to 0$. For a small error probability, from (2) we have

$$2\mathcal{P}(a) \leq \mathcal{B}(a) \leq 2\sqrt{\mathcal{P}(a)}.$$

Now let $\zeta = \tau^2/4$. If $\mathcal{P}(a_\ell) \leq \zeta$, then $t_\ell \leq \tau$. Thus, as $\ell$ tends to infinity, $\mathcal{P}(a_\ell) \to 0$ and $a_\ell \to \Delta_\infty$. ∎

The result can immediately be extended to a multitude of subchannels and their convex hull.

## IV. CHANNEL DECOMPOSITION

In this section, we propose a channel decomposition method based on the pdf of error probability defined in Section II. We show that every symmetric channel with capacity $\mathcal{C}$ can be decomposed into a number of basis channels with the same capacity. By the result of Section III, a code that works on the basis channels with capacity $\mathcal{C}$ is expected to be suitable for all channels in the convex hull of those basis channels.

We seek an identical-capacity basis for a given capacity $\mathcal{C}$ and contend that the sought basis is in the form of

$$g_{x,y}(p) = \alpha(x,y)\Delta_x(p) + \bar{\alpha}(x,y)\Delta_y(p)$$

where $\alpha(x,y) \in [0,1]$ is a constant depending on the channel capacity and the location of mass points $x$ and $y$. Since the basis channel $g_{x,y}(p)$ must have capacity of $\mathcal{C}$, by defining $\mathcal{H} = 1 - \mathcal{C}$ we have $\alpha(x,y)h(x) + \bar{\alpha}(x,y)h(y) = \mathcal{H}$ which results in

$$\alpha(x,y) = \frac{h(y) - \mathcal{H}}{h(y) - h(x)}.$$

Let $\xi = h^{-1}(\mathcal{H})$ be the crossover probability of a single BSC with capacity $\mathcal{C}$ where function $h^{-1} : [0,1] \mapsto [0, \frac{1}{2}]$ is the inverse of the binary entropy function. By the capacity constraint, in order to have a nonnegative $\alpha(x,y)$, we must have $\min\{x,y\} \leq \xi$ and $\max\{x,y\} \geq \xi$. Without loss of generality we assume that $x \leq y$.

Now we prove that every symmetric channel with capacity $\mathcal{C}$ falls into the convex hull of the basis channels

$$\mathcal{G}(\mathcal{C}) = \{g_{x,y}(p) | (x,y) \in \mathcal{D}(\mathcal{C})\}$$

where $\mathcal{D}(\mathcal{C}) = \{(x,y) \in [0,\xi] \times [\xi, \frac{1}{2}]\}$. To this end, we find a two dimensional pdf $\varphi(x,y)$ which fully describes the channel according to

$$g(p) = \int_{\mathcal{D}(\mathcal{C})} \varphi(x,y) g_{x,y}(p) dxdy \tag{4}$$

where

$$\int_{\mathcal{D}(\mathcal{C})} \varphi(x,y) dxdy = 1.$$

The following theorem shows that for any symmetric channel, there exists a two dimensional pdf $\varphi(x,y)$. Before stating the theorem, it is important to note that if the given channel has a mass point at $p = \xi$, then for some $\delta \in [0,1]$, we can write its pdf as $g(p) = \delta\Delta_\xi(p) + \bar{\delta}r(p)$ and proceed with $r(p)$ which does not have any mass point at $p = \xi$.

*Theorem 2:* Given a capacity $\mathcal{C}$, every pdf of error probability $g(p)$ associated with a symmetric channel can be written as

$$g(p) = \int_{\mathcal{D}(\mathcal{C})} \varphi(x,y) g_{x,y}(p) dxdy,$$

for some pdf $\varphi(x,y)$ defined over $\mathcal{D}(\mathcal{C}) = \{(x,y) \in [0,\xi] \times [\xi, \frac{1}{2}]\}$ where $\xi = h^{-1}(\mathcal{H})$.

*Proof:* Defining

$$\varphi(x,y) = \frac{g_\ell(x) g_r(y)}{\int_\xi^{\frac{1}{2}} g_r(\tau) \alpha(x,\tau) \frac{\bar{\alpha}(x,y)}{\bar{\alpha}(x,\tau)} d\tau},$$

one can verify that $\varphi(x,y)$ is a valid pdf and (4) holds. More discussions on $\varphi(x,y)$ and detailed proof are provided in [17]. ∎

Without loss of generality and for practical reasons, we assume that $|\mathcal{Y}|$ is finite. Having a finite alphabet also means having a finite basis, because the number of channels that satisfy (4) is finite. Assuming that the quantization levels are such that there are $N_\ell$ levels less than $\xi$ and $N_r$ levels greater than $\xi$, the cardinality of basis channels set is $|\mathcal{G}(\mathcal{C})| = N_\ell \times N_r$.

## V. SIMULATION RESULTS

Given the channel capacity $\mathcal{C}$ and the number of quantization levels, we can determine the set of basis channels $\mathcal{G}(\mathcal{C})$ for which we can design a single LDPC code. According to Conjecture 1, this code is expected to exhibit good performance on all channels in the convex hull of basis channels, i.e., $\mathfrak{C}[\mathcal{G}(\mathcal{C})]$. For a given code, we define the universal threshold $\mathcal{C}_u^*$ as the minimum capacity on which convergence of the code is guaranteed. Using Conjecture 1 and Theorem 2, the universal threshold $\mathcal{C}_u^*$ can be found by a binary search.

For all the simulations in this section, we have used 6-bit resolution to quantize subchannels (i.e., $|\mathcal{Y}| = 64$ at the receiver) and a 9-bit sum-product decoder [18]. We follow the guidelines in the literature and consider only two consecutive degrees for $\rho(x)$. Better universal codes can be obtained by optimizing $\rho(x)$, but the performance loss due to this simplifying assumption is minor.

Fig. 2 depicts a bit error rate comparison between a rate 0.6 code optimized for an additive white Gaussian noise (AWGN) channel with a maximum variable node degree of 100 [9], and a rate 0.6 universal code with the suggested method having maximum variable node of 50. Despite the fact that the AWGN code performs slightly better on the AWGN channel, we can see that the universal code outperforms the AWGN code on other channels.

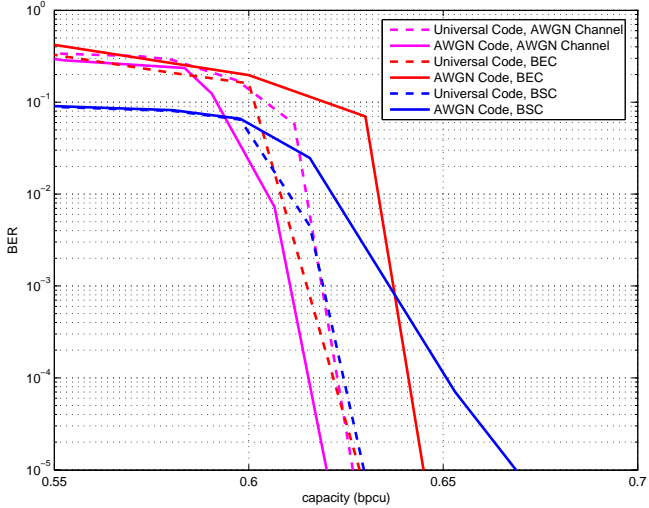

Fig. 2. Comparison between bit error rate of a rate 0.6 universal code and a rate 0.6 code designed specifically for AWGN on various channels. Both codes are randomly constructed and $100,000$ bits long.

The universal threshold for the AWGN code is $\mathcal{C}_u^* = 0.662$ while the universal threshold of the less complex universal code is $\mathcal{C}_u^* = 0.628$. Using density evolution, we tested both codes on $10,000$ randomly generated channels with the capacity of their universal threshold. The codes performed well on all channels.

It can be seen that the universal code has similar performance on all channels whereas the code that is designed for a specific channel only performs well on the channel for which it is designed. Degree distributions for the AWGN code taken from [9] are

$$\rho(x) = 0.5x^{12} + 0.5x^{13},$$
$$\lambda(x) = 0.1499x + 0.1621x^2 + 0.0224x^5 + 0.1764x^6$$
$$+ 0.0077x^7 + 0.1166x^{16} + 0.0307x^{27} + 0.0319x^{28}$$
$$+ 0.0438x^{30} + 0.0278x^{31} + 0.0048x^{42} + 0.2258x^{99}$$

while for the universal code are

$$\rho(x) = 0.5806x^{11} + 0.4194x^{12},$$
$$\lambda(x) = 0.1689x + 0.1924x^2 + 0.0604x^5 + 0.2069x^6$$
$$+ 0.0763x^{10} + 0.0457x^{29} + 0.2495x^{49}.$$

## VI. Conclusion

Design of universal LDPC codes over symmetric channels was discussed. We conjectured that, codes that converge on a set of equal-capacity channels, also converge on the convex hull of those channels. Therefore, we expect codes that are designed using this result to exhibit stronger universal properties than codes designed for specific channels. We also proved the stability of decoder in the convex hull of a set of equal-capacity channels. In other words, for a code which converges on a number of identical-capacity channels and on any channel in the convex hull of these channels, if the decoder gets close to the perfect decoding, then it will converge.

We proposed a channel decomposition technique which allowed for spanning any given channel with capacity $\mathcal{C}$ on a number of basis channels with identical capacity. Then, we designed codes for those basis channels. As expected, the codes designed following this method exhibited strong universal performance. Specifically, in comparison with existing LDPC codes designed for a given channel, significant performance gain was obtained when transmission took place over various channels of equal capacity. Defining a universal threshold for a code, we observed that our codes have better universal threshold compared to codes designed for specific channels.